\documentclass[12pt,epsf]{article}
\usepackage{epsfig}

\newcommand{\be}{\begin{equation}}
\newcommand{\ee}{\end{equation}}
\newcommand{\bea}{\begin{eqnarray}}
\newcommand{\eea}{\end{eqnarray}}

\begin{document}

\begin{flushright}
% SLAC-PUB-10916\\
% Feb 2005\\
\end{flushright}

\bigskip\bigskip
\begin{center}
{\bf\large CMB Fluctuation Amplitude from
Dark Energy Partitions\footnote{\baselineskip=12pt Work
supported by Department of Energy contract DE--AC03--76SF00515.}
\footnote{Work
supported by Department of Energy contract W-7405-Eng-48}}
\end{center}

\begin{center}
James V. Lindesay\footnote{Permanent address, Physics Department, Howard University,
Washington, D.C. 20059}, jlslac@slac.stanford.edu\\
H. Pierre Noyes, noyes@slac.stanford.edu \\
Stanford Linear Accelerator Center MS 81,
Stanford University \\
2575 Sand Hill Road, Menlo Park CA 94025\\
E.D. Jones, jones37@llnl.gov\\
Lawrence Livermore National Laboratory, Livermore, CA.   94551
\end{center}

{\small 
It is assumed that the dark energy observed today is
frozen as a result of a phase transition involving the
source of that energy.  Postulating that the dark energy de-coherence
which results from this phase transition drives statistical variations
in the energy density specifies a class of cosmological models
in which the cosmic microwave background (CMB) fluctuation
amplitude at last scattering is approximately $10^{-5}$. }


\bigskip
\setcounter{equation}{0}

% \textbf{I.  Introduction}
In a previous paper (reference \cite{JVLHPN2}, equation 14),
it has been shown that the fluctuation amplitude of the 
Cosmic Microwave Background (CMB) can be theoretically
estimated to be of the order $10^{-5}$, in agreement with
observational evidence, using a minimal set of assumptions.
Although the scale parameter of the cosmological expansion at the time
of dark energy de-coherence (as defined in reference \cite{JVLHPN2})
enters the calculation, the final
result depends only on the normalized dark energy density
$\Omega_\Lambda$, the red shift at last scattering $z_{LS}$,
and the red shift at the time when the energy density of the
non-relativistic pressure-less matter observed today was
equal to the radiation energy density $z_{eq}$.  The
observation that the final result was independent of any
specifics at the time of de-coherence suggested to the
authors of this paper that this result is more general than
the mechanism suggested in reference \cite{JVLHPN2}. 
This paper examines the generality of the previous result,
and presents a plausible justification for the size of
the fluctuations at the time of last scattering.
A tentative conclusion is
that any \emph{phenomenological} theory that accepts
the ``cosmological constant" $\Lambda$, or the dark
energy density $\rho_\Lambda={\Lambda \over 8 \pi G_N}$, 
or the deSitter scale radius $R_\Lambda=\sqrt{{3 \over \Lambda}}$, 
as a reasonable way to fit the
observational data indicating a flat, accelerating universe
over the relevant range of red shifts,
and which also  assumes some sort of phase transition that
decouples the residual dark energy from the subsequent
dynamics of the energy density, will fall within a class
of theories all of which fit the magnitude of the observed
fluctuations. 

% \textbf{II.  Motivation of form for amplitude of fluctuations}
A likely candidate for the dark energy is some form of vacuum
energy. 
Vacuum energy can often be thought of as resulting from
the zero-point motions of the sources
(independent of the couplings involved)\cite{Lifshitz}. 
As these motions de-cohere and become localized,
the deviations from uniformity are expected to appear as
fluctuations in the cosmological energy density. 
Since these motions are inherently
a quantum effect, one expects the fluctuations to exhibit the
space-like correlations consistent with a quantum phenomenon.

A weakly interacting sea of the quantum fluctuations due to
zero point motions should exhibit local statistical variations in
the energy.  For a sufficiently well defined phase state,
one should be able to use simple invariant counting
arguments to quantify these variations.  If the zero-point motions
of the sources have a statistical weight $\Omega(E_A)$ associated
with a partition $A$ having energy $E_A$ while holding total energy
fixed, then the probability of such a partitioning is given by
\be
P(E_A) \: = \: {\Omega(E_A) \over \Omega_{tot}} \: = \:
{\Omega_A (E_A) \, \Omega_{\bar{A}} (E_{tot} - E_A) 
\over \Omega_{tot}},
\ee
where $\bar{A}$ represents everything external to the
(statistically independent) partition $A$.
If one requires that the most likely configuration of energy
partitions results when (the log of) this probability is maximized
with respect to the energy partition $E_A$,
this distribution has uniform dark energy when
\be
{1 \over E_\Lambda ^A } \: \equiv  \: { d \over dE_A } log \Omega_A (E_A)
\quad \quad , \quad \quad E_\Lambda ^A = E_\Lambda ^ {\bar{A}}
\equiv E_\Lambda .
\ee
Here $E_\Lambda$ is an intensive energy (chemical potential) associated with the
statistical reservoir and boundary conditions. 
This result is analogous to the zeroth law of thermodynamics.
Similarly, using arguments analogous to those
used to establish the second law of
thermodynamics, $log \Omega (E)$ is expected to be a non-decreasing
function for previously isolated systems placed in mutual contact.

If one next considers a ``small" partition $A$ for which the reservoir
$\bar{A}$ has energy $E_{tot}-E_A$, one can examine the (log of the) lowest order
fluctuations from uniformity for the reservoir to show that
\be
\Omega_{\bar{A}} (E_{tot}-E_A) \: \cong \: \Omega_{\bar{A}} (E_{tot}) \,
e^{-E_A / E_\Lambda},
\ee
thus defining a probability distribution in terms of the energy
E of the "small" partition given by
\be
P(E) \: = \: {e^{-E/E_\Lambda} \over \sum_{E'} e^{-E' /E_\Lambda} }.
\ee
For such an ensemble, the energy spread from the mean is given by
\be
<(\delta E)^2> \: = \: E_\Lambda ^2 {d \over dE_\Lambda} <E>.
\ee
A typical equation of state will connect the extensive variable $<E>$
to another extensive variable that counts the available degrees
of freedom $N_{DoF}$.
On dimensional grounds, the terms in a typical equation of state
which depend on $E_\Lambda$ should take
the general form $<E>=N_{DoF} {(E_\Lambda )^{a} \over \epsilon^{a-1}}$,
where $\epsilon$ is a constant with dimensions of energy.  The
expected fluctuations are then given by
\be
{<(\delta E)^2> \over <E>^2} 
\: = \: {a \over N_{DoF}} \left ( {\epsilon \over E_\Lambda} \right ) ^{a-1}
\: = \: a \, {E_\Lambda \over <E>} \, .
\label{delE}
\ee
As seen from the second form in equation \ref{delE}, the
dimensionless fluctuations are typically of the order of ${1 \over N_{DoF}}$,
i.e. inversely proportional to a dimensionless extensive parameter
that typifies the scale of the system. 
In terms of the densities, one can directly write
${<(\delta E)^2> \over <E>^2} \: = \: {<(\delta \rho)^2> \over \rho^2}
\sim  {\rho_\Lambda \over \rho}$.

Therefore, the energy available for fluctuations in the two point correlation
function is expected to be given by the cosmological dark energy,
in a manner similar to the way that background thermal energy $k_B T$
drives the fluctuations of thermal systems.  This means that the
amplitude of relative fluctuations $\delta \rho / \rho$
is expected to be of the order 
\be
\Delta_{PT} \: \equiv  \: \left (  {\rho_\Lambda \over \rho_{PT} } \right ) ^{1/2}  
\label{DelPT}
\ee 
where $\rho_{PT}$ is the cosmological energy density
at the time of the phase transition that decouples the dark energy.
The scale dependence of this
form will be explored in the remainder of this letter.

% \textbf{III.  Evolution from radiation domination regime}
Next, such a phase transition which occurs during the epoch
of radiation domination will be considered. 
Using the densities at radiation-matter equality $\rho_M (z_{eq})=\rho_{rad} (z_{eq})$,
one can extrapolate back to the phase transition period to determine the redshift at that time. 
The (non-relativistic) baryon-electron plasma is expected to scale using
$\rho_M (z)=\rho_{Mo} (1+z)^3$ until it is negligible, whereas, the radiation
scales during the early expansion using $\rho_{rad} (z)=\rho_{PT} \left ( {1+z \over 1+z_{PT}}
\right )^4$.
Ignoring threshold effects (which give small corrections near particle thresholds while they
are non-relativistic), this gives
\be
1+z_{PT} \: = \: \left [ {\rho_{PT} \over \rho_{Mo}} (1+z_{eq})  \right ]^{1 \over 4} .
\label{zPT}
\ee
Here, $\rho_{Mo}$ is the present pressure-less mass density.

For adiabatic perturbations (those that fractionally perturb the number
densities of photons and matter equally),
the matter density fluctuations grow according to\cite{PDG}
\be
\Delta \: = \: \left \{
\begin{array}{cc}
\Delta_{PT}  \left ( {R(t) \over R_{PT}} \right ) ^2  &
radiation-dominated \\
\Delta_{eq} \left ( {R(t) \over R_{eq}} \right ) &
matter-dominated
\end{array}
\right . .
\ee
This allows an accurate estimation for the scale of fluctuations
at last scattering in terms of fluctuations during the
phase transition given by
\be
\Delta_{LS} \: = \: \left( {R_{LS} \over R_{eq}}  \right)
\left( {R_{eq} \over R_{PT}}  \right) ^2 \Delta_{PT} \: = \:
{ (1+z_{PT}) ^2 \over (1+z_{eq}) (1+ z_{LS}) } \Delta_{PT} .
\label{DelLS}
\ee

Using equations
\ref{zPT}, \ref{DelLS}, and \ref{DelPT}, this amplitude at last scattering
is given by 
\be
\Delta_{LS} \: = \: { (1+z_{PT}) ^2 \over (1+z_{eq}) (1+ z_{LS}) }
\left (  {\rho_\Lambda \over \rho_{PT} } \right ) ^{1/2} \: \cong  \: {1 \over 1 + z_{LS}} \sqrt{{\Omega_{\Lambda o}
\over (1-\Omega_{\Lambda o}) (1+z_{eq})}} \cong 2.5 \times 10^{-5}, 
\ee 
where a spatially flat cosmology and radiation domination has been assumed.
The values taken for the phenomenological parameters
are given by $\Omega_{\Lambda o}\cong 0.73$, 
$z_{eq} \cong 3500$, and $z_{LS} \cong 1100$.  
This estimate for a transition during the radiation
dominated regime is independent of the density during
the phase transition $\rho_{PT}$, and is of the order observed for
the fluctuations in the CMB (see \cite{PDG} section 23.2 page 221).
It is also in line with those argued by other
authors\cite{Padmanabhan} and papers\cite{ANPA}. Fluctuations in the CMB at last
scattering of this order are consistent with the currently
observed clustering of galaxies.

% \textbf{IV.  Evolution from dust domination regime}
If the phase transition were to occur during the epoch of pressure-less
matter domination, $z_{eq}>z_{PT}>z_{LS}$, the fluctuation amplitude
will be seen to demonstrate a weak dependence on the time of the phase transition. 
The acoustic wave has coherent phase information that is transmitted
to the CMB at last scattering.  There must have been a significant enough
passage of time from the creation of the acoustic wave to the time of
last scattering such that peaks and troughs of the various modes
should be present at $\delta t > {\lambda \over v_s}$, where
$\lambda$ is the distance scale of the longest wavelength (sound horizon), and
$v_s \sim c/\sqrt{3}$ is the speed of the acoustic wave.
Generally, if the phase transition occurs while the energy
density is dominated by dark matter/plasma,
then the amplitude satisfies
\be
\begin{array}{c}
\sqrt{{\rho_\Lambda \over \rho_{PT}}} \: = \:
\sqrt{{\Omega_{\Lambda o} \over \Omega_{rad \, o} (1+z_{PT})^4 +
\Omega_{M o} (1+z_{PT})^3 + \Omega_{\Lambda o} }} \\ \\ \: = \:
\sqrt{{\Omega_{\Lambda o} \over ( 1 - \Omega_{\Lambda o} )
(1+z_{PT})^3 \left( {1+z_{eq} \over 2 + z_{eq}} \right )
\left (
1+ {1+z_{PT} \over 1+z_{eq}}
\right ) + \Omega_{\Lambda o} } } \: .
\end{array}
\ee
This gives an amplitude at last scattering of the order
\be
\Delta_{LS}  \: \cong \:
\left ( {1+z_{PT} \over 1+z_{LS}}  \right )
\sqrt{{\Omega_{\Lambda o} \over 
(1-\Omega_{\Lambda o}) (1+z_{PT})^3 
\left (
1+ {1+z_{PT} \over 1+z_{eq}}
\right ) }} \: ,
\ee
which varies from $2 \times 10^{-5}$ if the phase transition
occurs at radiation-matter equality, to $4 \times 10^{-5}$ if it
occurs at last scattering.

% \textbf{V.  Cold Dark Matter Mass Limits}
As an example of a phase transition while the involved particles are
non-relativistic such as has been discussed,
consider cold dark bosonic matter made
up of particles of mass $m$.
For non-relativistic bosonic dark matter, the
relationship between number density and critical
density for Bose condensation for a free Bose gas is given by\cite{Bose}
\be
{N \over V} \: = \: {\zeta({3 \over 2}) \Gamma({3 \over 2}) \over (2 \pi)^2 \hbar ^3}
\left ( 2 m k_B T_{crit} \right ) ^{3/2}.
\ee
Since the dynamics is assumed non-relativistic,
$\rho_m \cong {N \over V}m c^2$, giving the following
requirement for a macroscopic quantum system
made up of Bose condensed cold dark matter:
\be
\left( m c^2  \right )^{5/2} \: \cong  \:
{\rho_m \over \left( 2 k_B T_{crit} \right )^{3/2}}
{  (2 \pi)^2 ( \hbar c )^3 \over \zeta({3 \over 2}) \Gamma({3 \over 2}) }.
\label{CDMlimit}
\ee
In order for the macroscopic space-like quantum coherent
state to persist, the ambient temperature must be less than
the critical temperature.  If the phase transition occurs while
the dark matter is cold ($m>k_B T_{PT}$), its density can be assumed to
depend on the redshift by $\rho_m = \rho_{mo} (1+z)^3$.  The
temperature of the photon gas is expected to likewise scale with the
redshift when appropriate pair creation threshold affects are properly
incorporated in the manner 
$T_\gamma (z) \approx  T_{\gamma o} (1+z) (g(0) / g(z))^{1/4}$,
where $g(z)$ counts number of low mass thermal degrees of freedom
available at redshift $z$.  Substitution 
of this photon temperature as the limiting critical temperature
into equation \ref{CDMlimit}
gives
\be
\left( m c^2  \right )^{5/2} \: < \:
(1+z)^{3/2} \left ( {g(z) \over g(0)} \right )^{3/8}
{\rho_{m o} \over \left( 2 k_B T_{\gamma o} \right )^{3/2}}
{  (2 \pi)^2 \hbar ^3 \over \zeta({3 \over 2}) \Gamma({3 \over 2}) }.
\ee
Thus, the upper limit on a condensate mass roughly satisfies
\be
mc^2 \: < \: (1+z)^{3/5} \left ( {g(z) \over g(0)} \right )^{3/20} \, \times
(1.2 \times 10^{-11} GeV).
\ee
If the transition occurs as late as last scattering, 
a scalar Bose condensate model for cold dark
matter would require a particle mass below
$0.8 eV$.  However, if the transition occurs during an earlier epoch,
the mass of the condensate
particles can be considerably larger.

% \textbf{VI.  Conclusions}
In conclusion, it has been shown that if a general phase transition
which freezes the cosmological effects of the dark energy
(which thereafter can be represented as a cosmological constant)
occurs sufficiently prior to last scattering, statistical
fluctuations driven by the dark energy produce
density perturbations of a magnitude that will
evolve to be of the order $3 \times 10^{-5}$ at last
scattering.  It should be noted that this result
is independent of the details of the
mechanism of phase transition.
\bigskip

\end{document}